\documentstyle[12pt]{article}   
\begin{document}

\centerline{{\Large\bf
 Collectivity, Phase Transitions and}} 
\centerline{{\Large\bf Exceptional Points in Open Quantum Systems}}   
\bigskip   
\centerline{{\sc W.D. Heiss$^{1,2}$,   
M. M\"uller$^2$ and I. Rotter$^{2,3}$}}   

\centerline{{\sl $^{1}$Department of Physics, University of the   
Witwatersrand}}   
\centerline{{\sl PO Wits 2050, Johannesburg, South Africa}}   
\centerline{{\sl $^{2}$ Max-Planck-Institut f\"ur Physik   Komplexer   
Systeme}}   
\centerline{{\sl 01187 Dresden, Germany}} 
\centerline{{\sl $^3$ Technische Universit\"at Dresden, 
Institut f\"ur Theoretische Physik}}  
\centerline{{\sl D-01062 Dresden, 
Germany}} 
 
\vspace{1cm}

05.70.Fh, 03.80.+r, 64.60.-i, 02.30.Dk\\[.2cm]

\begin{abstract}   
Phase transitions in open quantum systems, which are associated with the
formation of collective states of a large width and of trapped states with
rather small widths, are related to exceptional points of the Hamiltonian.
Exceptional points are the singularities of the spectrum and eigenfunctions,
when they are considered as functions of a coupling parameter.
In the present paper this parameter is the coupling strength to the
continuum. It is shown that the positions of the exceptional points (their
accumulation point in the thermodynamical limit) depend on the particular
type and energy dependence of the coupling to the continuum in the same way
as the transition point of the corresponding phase transition.
\end{abstract}   
   
\newpage
   
\section{Introduction}

Recently, mechanisms of restructuring of quantum systems   
are discussed with renewed interest 
\cite{trap} - \cite{PT}.
Conditions for the formation of   
collective states on the one hand and of quantum chaos on the other are    
being studied.   
Among the oldest examples of the former is the schematic model explaining in  
simple terms the origin of giant dipole resonances in nuclei \cite{schemat}. The  
Hamiltonian is of the form $H=H_0 + H_1 =
H_0 + D D^T$ where $H_0$ is the Hamiltonian  
of the unperturbed system  and $D D^T$ describes the factorized dipole-dipole  
residual interaction.   
The rank of $H_0$ is $N$ where $N$ is the number of unperturbed states   
considered in a certain energy interval, while the rank of $D D^T$ is 1.    
If the modulus of the average matrix elements of the vector $D$ is 
sufficiently large
($\bar D \gg d_0$, where $\bar D$ is the average matrix element of $D$ and
$d_0$ is the mean level distance of the eigenvalues of $H_0$), one of 
the eigenstates of $H$ is shifted considerably in energy. Since its
 eigenvector has contributions from almost all basis    
states, it is a collective state.   
This type of collectivity is called  {\it internal} collectivity \cite{sorosamu}.   
\\   
   
Besides the internal collectivity, there exists the so-called external 
collectivity of resonance states \cite{sorosamu}.
It appears at high level density since the discrete states   
are in general embedded in the continuum and coupled to each other via the   
continuum. The Hamiltonian of the  
open quantum system is, to a good approximation, given by  
 $H_0 - i V V^T$ where $V V^T$ contains the coupling matrix elements  
$V_j^c$ between  
the discrete states $j$ and the decay channels $c$. The rank   
of $VV^T$ is equal to the number $K$ of open decay channels. It is  
usually much smaller than the rank $N$ of $H_0$.   
If the matrix elements  of $VV^T$ 
 are sufficiently large  ($\bar V \gg d_0)$,
 $K$ collective states are formed. 
 They are distinguished  from the $N-K$ non-collective states by their   
large widths or short lifetimes \cite{trap} - \cite{PT}.
The wavefunctions of these $K$ collective resonance states  
are again characterized by a large number of components of the basis states.   
 \\ 
  
There is a basic difference between internal and external collectivity. 
Internal collectivity deals with a self-adjoint Hamiltonian, where the 
coupling 
parameter is real, while external collectivity is described by a 
 purely imaginary coupling strength.  
The particular type of collectivity is therefore expressed by a real spectrum 
in the first case, whereas in the second case the dissipative character of  
the open system is reflected by imaginary parts of the 
eigenenergies, the physical widths.
Formally, the formation of a collective state is, however, the same in 
both cases.
It is expressed by the mixing of the eigenfunctions of the full Hamiltonian
with respect to the eigenfunctions of $H_0$. \\

The  case  ${\rm rank}(H_0)={\rm rank}(H_1)$ is not 
necessarily connected
with the appearance of collective states. It
has been widely discussed in the literature \cite{hesa}.  
Conditions have been given for the occurrence of chaotic behaviour  
and of some sort of phase transitions \cite{wdh}.  
The mechanism of such restructuring  
and possible chaotic behaviour in the transitional region  
($\Lambda \sim \Lambda _{c}$) have been explained in terms of the exceptional  
points (EPs) of the problem $H_0+\Lambda H_1$. \\

The EPs are the  only singularities in the complex $\Lambda -$plane of  
the eigenvalues  $E_k(\Lambda )$. Their positions are fixed by the 
choice of $H_0$ and $H_1$ only. As a consequence,   
the distribution of the EPs is characteristic for any  
particular Hamiltonian of the form $H_0+\Lambda H_1$, once $H_0$ and $H_1$ are 
given. \\
 
It is therefore natural to discuss   
the EPs as they determine important properties of the spectrum such   
as the statistical properties, the type and locations of avoided level   
crossings, the softness of the spectrum and the ranges of $\Lambda -$values, 
where   
special features of the spectrum occur. In particular, the distribution   
and frequency of the  EPs give indications about transitional   
regions and possible occurrence of chaotic behaviour. \\   
   
The studies of the EPs have, in the papers quoted,   
been dealing with closed systems, where the influence of decay channels   
in the continuum has not been considered, that is $H$ is selfadjoint   
and the eigenstates are discrete. As stated above we are often faced with    
dissipative systems, where the coupling to the continuum and decay channels is 
expressed by complex eigenenergies which give rise to   
finite life times of resonance states.   
The complex eigenvalues ${\cal E} _k=E_k-i/2\Gamma_k$ are interpreted  
as resonance states at the energies $E_k$ with the decay widths $\Gamma_k$    
which are inverse life times.  
\\   
   
While there is a great mathematical   
similarity in the treatment of the respective self-adjoint and   
dissipative Hamiltonians, the physical findings deserve particular attention. 
Of special interest will be the circumstances, under which the formation   
of collective states can be understood as a phase transition.   
We generalize the results obtained in \cite{wdh} to open quantum    
mechanical systems and expand the investigations of \cite{PT}   
by taking into account the effects and properties of the EPs. 
In all cases, we will restrict ourselves to the one-channel case 
in which 
the phase transition as well as the collectivity of one of the states 
are well pronounced.
\\

Generally, a phase transition is 
a substantial restructuring of the system taking place at a finite
(critical) value $\Lambda _c$ of a certain control parameter $\Lambda$. 
This restructuring is a collective phenomenon,
ranging over all scales inherent in the system.
The nature of the reorganization process is characterized by the 
behaviour of  an order parameter being a function of 
$\Lambda$. 
In our case of an open quantum system the order parameter is 
$\Gamma_0/N$ where $\Gamma_0$ is the width of the collective mode
(in the one channel case) and $N$ is the total number of states.
Its first derivative with respect to $\Lambda$ shows a finite discontinuity at
the critical value $\Lambda _c$ corresponding to a second-order
phase transition. 
All characteristic features of the phase transition 
emerge already at  finite values of $N$ despite  the fact that
the strict thermodynamical definition is possible only for $N \to \infty$. 
Precise conditions for its occurence are derived in \cite{PT}.\\

The phase transition is connected with the appearance of a state 
whose external collectivity is of a global 
nature. Its wave function carries contributions from {\it all} eigenstates 
of $H_0$. This global collectivity must be distinguished
from the local (external) collectivity which appears 
when the resonance states are coupled strongly to the continuum but the
conditions for a phase transition are not fulfilled.         
The wave function of this collective state carries contributions only 
from a restricted number of basis states, whose 
unperturbed energies are overlapped by the width of the collective 
state \cite{PT}. \\

In Section 2 the basics of the EPs and  specific features relating to 
open quantum systems are reviewed.   
A simple example using two resonances coupled to one decay channel  
illustrates the connection between the main results known from  
the study of open quantum mechanical systems \cite{mudiisro} 
and the aspect of the EPs. 
In Section 3 the restructuring of the system is put into context with   
several systems of a different nature.
As a beginning, the simplest 
case of a picket fence model is considered.
 Characteristic patterns of the EPs relating to phase transitions are
established. Next, the influence of random perturbations within
the system is investigated. Finally we discuss the   
effect of a more general level density distribution of the unperturbed levels 
and that of the coupling vector $V$.  
A short summary and outlook is given in Section 4.\\

\section{Exceptional points and features of open \newline quantum systems}   
   
The $N\times N$ matrix problem of the form $H_0+\Lambda H_1$ has eigenvalues   
$E_k(\Lambda ),\,k=1,\ldots ,N$ which are obtained by the secular equation   
\begin{equation}   
\det (H_0+\Lambda H_1-E)=0.  \label{sec1}   
\end{equation}   
We here assume that $H_0$ and $H_1$ are real and symmetric. The   
EPs \cite{hesa} are characterized by the coalescence of any two pairs of   
eigenvalues.  They are fixed by a particular choice of $H_0$ and $H_1$.   
If we exclude a genuine degeneracy of eigenvalues for real values   
$\Lambda $, such coalescence will happen only for complex values of    
the coupling parameter and hence for complex eigenvalues. 
Accordingly, the EPs are   
determined by the simultaneous solutions of Eq.(\ref{sec1}) and of   
\begin{equation}   
{{\rm d}\over {\rm d}E }\det (H_0+\Lambda H_1-E)=0  \label{sec2}   
\end{equation}   
where the second equation ensures that two eigenvalues coincide.   
Eqs.(\ref{sec1}) and (\ref{sec2}) are polynomials in $E$ and $\Lambda $ of order   
$N$ and $N-1$, respectively. They can be combined into a single polynomial   
in $\Lambda $ of order $N(N-1)$, the resultant of Eq.(\ref{sec1}), by eliminating   
the variable $E$. The resultant has real coefficients, hence the    
EPs occur in $N(N-1)/2$ complex conjugate pairs. They are the only   
singularitites which the eigenvalues $E_k(\Lambda )$ can have as functions of   
$\Lambda $. In fact, they are the square root branch points of one analytic   
function which has $N$ Riemann sheets, where the values on each sheet   
are just the real eigenvalues $E_k(\Lambda )$ for real values $\Lambda $. All   
$N$ eigenvalues are therefore   
analytically connected with each other via the square root branch points   
in the complex $\Lambda -$plane. \\   
   
The behaviour of the eigenfunctions when continued to the EPs    
deserves special interest.    
Since they are lying in the complex $\Lambda$-plane, the operator  
$H_0+\Lambda H_1$ is no longer self-adjoint at $\Lambda =\Lambda _{\rm EP}$ with   
$\Lambda _{{\rm EP}}$ denoting a (complex) EP. Therefore we cannot expect two   
linearly independent   
eigenvectors even though two (complex) eigenvalues $E_i$ and $E_k$ coalesce.   
The eigenbasis of $H_0+\Lambda _{{\rm EP}} H_1$ is no longer orthogonal as it is   
the case for selfadjoint Hamiltonians.   
In fact, in contrast to the self-adjoint case, where a twofold degeneracy   
always implies a two-dimensional eigenspace, an EP is   
characterized by the fact, that the rank of the associated   
matrix $H_0+\Lambda _{{\rm EP}}H_1$   
drops  by one. In other words, there is no two-dimensional subspace   
associated with the coalescence of two eigenvalues. We rather encounter a   
confluence of the two eigenvectors $\psi _l(\Lambda )$ and $\psi _k(\Lambda )$ at   
$\Lambda =\Lambda _{{\rm EP}}$. Moreover, for a general complex value $\Lambda $ not  
coinciding  with an EP, we may choose a biorthogonal system such that   
\begin{equation}   
\langle \tilde \psi _l(\Lambda )|\psi _k(\Lambda)\rangle =\delta _{lk},   
\label{orth}   
\end{equation}   
where $\tilde \psi _i(\Lambda )$ and $\psi _i(\Lambda )$ are the left and the right
eigenvectors, respectively. We note, that 
the transformation $O$ diagonalizing a symmetric complex Hamiltonian $H$
 is complex orthogonal ($OO^T=O^TO=1$) but not unitary. Therefore,
the scalar product of the two left or right eigenvectors obeys   
$\langle\tilde\psi _k(\Lambda )|\tilde\psi _k(\Lambda)\rangle=  
\langle\psi _k(\Lambda )|\psi _k(\Lambda)\rangle \ge 1$   
for any $k$. A problem occurs when the normalization  
condition given by Eq.(\ref{orth}) is considered for $\Lambda \to \Lambda _{{\rm EP}}$: 
with $E_l(\Lambda _{{\rm EP}})=E_k(\Lambda _{{\rm EP}})$ it is  
$\psi _l(\Lambda _{{\rm EP}})=\psi _k(\Lambda _{{\rm EP}})$. This means, in view of 
Eq.(\ref{orth}), that $\psi _k(\Lambda _{{\rm EP}})$ cannot be normalized at 
$\Lambda =\Lambda _{{\rm EP}}$, since now the   
orthogonality conflicts with the normalization requirement. Usually,   
Eq.(\ref{orth}) is globally enforced; as a consequence, the two states 
$\psi _l(\Lambda )$ and $\psi _k(\Lambda )$ not only coincide for 
$\Lambda \to \Lambda _{{\rm EP}}$ but   they blow up, that is    
$\lim _{\Lambda \to \Lambda _{{\rm EP}}}\langle\psi _k(\Lambda )|\psi _k(\Lambda) \rangle   
\to \infty $ \cite{mudiisro}.\\

The physical significance of the EPs lies in their relation  
to avoided level crossing. In particular, in a number of examples it has been  
demonstrated that in the region of the real values of $\Lambda $, where a high   
density of EPs occurs, the typical statistical characteristics of the  
spectrum ascribed to quantum chaos prevail \cite{hesa,kohe}.\\  
  
Turning to open quantum systems 
\begin{eqnarray}
H=H_0- i \Lambda VV^{\dagger } 
\label{eq:heff}
\end{eqnarray}
we are faced with a Hamiltonian which has in general complex eigenvalues   
${\cal E} _k=E_k-i/2\Gamma_k$ as $\Lambda $ is complex,
\begin{eqnarray}
\Lambda = \lambda e^{i \varphi} \; .
\label{eq:lam}
\end{eqnarray}
(The  factor $i$ in front of the coupling term $\Lambda$
in Eq.~(\ref{eq:heff})  is used traditionally in this context).
The question then arises, whether and under which conditions two 
(or more) eigenvalues  coincide in the complex energy plane
 and how such a crossing depends on the    
coupling strength $\Lambda$. The answer to these questions depends on the   
manner by which the EPs are fixed by the operators $H_0$ and $V$. \\   
   
For illustration let us consider the simple example of two    
resonances, which are coupled to one   
open decay channel. The Hamiltonian matrix for this system can be written    
in the eigenbasis of $H_0$ as   
\begin{eqnarray}   
H &=& \pmatrix{\epsilon _1 & 0 \cr 0 & \epsilon _2 }   
- i \Lambda \pmatrix{\cos ^2 \omega &   
\cos \omega \; \sin \omega \cr \cos \omega \; \sin \omega & \sin ^2 \omega }   
\label{twod}   \cr  & & \cr 
 & \equiv & H_0 - i \Lambda \;  VV^{\dagger },   
\end{eqnarray}   
where we use $H_1=VV^{\dagger }$.   
The relative coupling strength of the two resonance states to the
 continuum is    
determined by the angle $\omega $ of the vector $V=(\cos\omega,\sin\omega)$.   
The simplicity of the model provides  an   
analytic expression for the two EPs, {\it viz.}   
\begin{equation}   
\Lambda _{{\rm EP}}=i(\epsilon _2-\epsilon _1)e^{\pm 2i \omega } 
\label{exslt}   
\end{equation} 
which are the zeros of the square root in the expression for the 
eigenenergies, which read 
\begin{equation} 
{\cal E} _{1,2}={\epsilon _1 +\epsilon _2-i\Lambda \over 2} \pm 
{1\over 2}\sqrt{(\epsilon _1-\epsilon _2)^2- 
2i\Lambda (\epsilon _1-\epsilon _2)\cos 2\omega +(i\Lambda )^2 }.   
\label{eig2} 
\end{equation} 
The example nicely demonstrates that the EPs depend {\it only} on $H_0$   
and $H_1$. Under variation of the angle $\omega$ the EPs are    
moving on a circle with    
radius $\epsilon _2-\epsilon _1$ in the complex $\Lambda -$plane. What    
determines the EPs are the energies of the unperturbed states 
(that is $H_0$)    
and their relative coupling (that is $H_1$), and {\it not} the   
phase $\varphi $ of the  coupling $\Lambda =\lambda e^{i\varphi }$.  \\ 
 
Yet, the complex eigenvalues ${\cal E}_{1,2}$   obviously {\it do} depend on    
$\Lambda $,  that is on its modulus $\lambda $ and on the angle $\varphi $.   
Variation of $\lambda $ invokes trajectories in the complex energy plane. The    
contours of the trajectories depend therefore on both, the phase $\varphi $    
and  the relative coupling controlled by $\omega $.   
For $\omega=45^0$ and $\varphi =\varphi _{{\rm EP}}=0^0$ the two trajectories 
will cross one EP   
at $\lambda = \lambda _{{\rm EP}} = \epsilon _2-\epsilon _1$. 
At this critical value of the    
coupling strength the two eigenvalues coalesce. 
The dimension of the eigenspace of    
the Hamiltonian is reduced to one including the consequences for the    
eigenvectors mentioned above. 
 We stress, however,  that the crossing of trajectories should not   
be confused with the familiar degeneracy for self-adjoint Hamiltonians.   
\\ 
 
For  $\varphi \ne \varphi _{\rm EP} $ the two    
trajectories repel each other within a certain {\it finite} distance in   
the complex plane \cite{mudiisro}. This is a generalization of avoided level 
crossing for real  eigenvalues of a hermitian Hamilton operator. 
Only if $\varphi $ is    
properly tuned to the value of $\omega $ can the two energy trajectories   
genuinely cross each other. 
A nice illustration of this simple example for different values of $\omega$    
and $\varphi$ can be found in \cite{pegoro}. 
 \\   
   
What becomes obvious in the $2\times 2$ matrix model can be generalized to an 
arbitrary $N\times N$ situation, i.e. $N$ resonance states coupled  
to one common decay channel. 
With increasing $\lambda $ one eigenvalue trajectory always    
drifts further into the complex plane,    
while the others are bending back towards the real axis    
after they have repelled with the collective state whose width 
always increases. This happens irrespective   
of $\varphi $. Physically it means that one of the resonance 
states  takes almost    
all of the transition strength by trapping the others which then 
become long-lived \cite{mudiisro,pegoro}.\\ 
 
 This is understood by the   
rank  one of the coupling matrix. At large values of  
$\lambda $ the first part $H_0$ of the Hamiltonian is    
a small perturbation. Therefore, the total matrix is essentially turned into
an operator of rank 1. In other words, there is only one nonzero eigenvalue, 
and the widths of the zero eigenvalues have to vanish. In    
general, with $N$ resonance states and $K<N$ open decay channels, 
there appear    
$K$ fast decaying states and $N-K$ states which are virtually stable. 
Various examples of the many channel -- many resonance case can be found    
in \cite{trap} - \cite{PT}. \\ 
 
The low rank of $H_1$ has a drastic effect on the total number of EPs. 
If $K$ is the rank of $H_1$,  
the number of EPs is $K(2N-K-1)$. The important point is  
the linear behaviour in $N$; only when $K$ attains the order of magnitude  
of $N$ is the quadratic behaviour retrieved.  
This finding is significant in that it indicates that chaotic behaviour  
cannot be {\it generated} by a low rank of $H_1$, there is  
simply an insufficient number of avoided level crossings. This is in line 
with an analysis of the level statistics of a Poisson ensemble coupled to a
continuum by a Gaussian coupling vector \cite{dirose,godimurose}, where, for 
large coupling 
strength, chaotic features of the system are the more pronounced
the larger the  number $K$ of decay channels. \\

\section{Results} 
   
We turn our attention to the critical region, where a restructuring of    
the system occurs. We address the relationship between the potential 
occurrence of a phase transition which is associated with the formation of a  
globally collectice state and the distribution of the EPs.   
Since the essential properties of the Hamiltonian must be reflected 
in the distribution   
of the EPs, we expect this relationship to exist. \\

\subsection{Simple example: picket fence}   
   
We consider the simple model with   
\begin{equation}   
H_0=\pmatrix{-{N-1\over 2}&0&\cdots&0 \cr   
         0&-{N-3\over 2}&\cdots&0 \cr   
         \vdots&&\ddots&\vdots \cr   
         0&\cdots&0&{N-1\over 2} \cr}     \label{h0}   
\end{equation}   
and the $N\times N$ matrix of rank one   
\begin{equation}   
VV^{\dagger }=\pmatrix{1&\cdots &1 \cr   
         \vdots&&\vdots \cr   
          1&\cdots &1 \cr} .    \label{h1}   
\end{equation}   
The Hamiltonian describes a picket fence spectrum, where all states are   
coupled equally to one decay channel. Such a system shows a phase transition
at $\Lambda
 = 1 / \pi$ \cite{PT}.
 Results   for finite $N$ can be obtained easily by   numerical means. 
In  the limit $N \to \infty$ the zeros of    
$\det (H_0-i\Lambda VV^{\dagger }-{\cal E})$ and   
$\sin (\pi {\cal E})-i\Lambda \cos (\pi {\cal E})$ coincide \cite{PT}. One   
finds an accumulation point of the EPs, which emerges in the limit   
and is found from the zeros of the resultant, which   
is obtained by eliminating the variable ${\cal E}$ from the set   
\begin{eqnarray}   
\sin (\pi {\cal E})+i\pi \Lambda \cos (\pi {\cal E})&=&0 \\   
\cos (\pi {\cal E})-i\pi \Lambda \sin (\pi {\cal E})&=&0.   
\end{eqnarray}   
These two equations are equivalent to Eqs.~(\ref{sec1}), (\ref{sec2}).
Their simultaneous solutions, the zeros of the
resultant (denoted by Rsl), are given by
\begin{equation}   
{\rm Rsl}(\lambda )=\sqrt{1+(i\pi \Lambda )^2}.
 \label{rsl}   
\end{equation}   
Obviously, this is no longer a polynomial. The important point 
of our finding   
is that {\it all} roots which occur in the complex $\Lambda -$plane for finite $N$ 
converge to  $\Lambda = \lambda =\pm 1/\pi$ in the limit $N\to \infty $.   
Negative values of $\lambda$ would lead to negative decay widths which have no
physical meaning. In the following, we restrict ourselves
 to positive $\lambda$
and $0^0 \le \varphi \le 90^0$, since all relations are symmetric
with respect to the replacement $\Lambda \to -\Lambda$ and $\Lambda \to \Lambda^*$ . 
Note also that the limit point, being a point of accumulation, is no   
longer a square root branch point for the energy spectrum but rather a   
logarithmic branch point.   
In fact, the analytic behaviour of the (infinitely many) energy levels   
${\cal E}_k(\Lambda )$ in the vicinity of $\Lambda =1/\pi $ is found by solving   
the secular equation explicitly for ${\cal E}$. The expression reads   
\begin{equation}   
{\cal E}(\Lambda )={1\over \pi }{\rm atan}(-i\Lambda \pi)=   
{i\over 2\pi }\ln{1-\Lambda \pi\over 1+\Lambda \pi} 
\label{enfct}   
\end{equation}   
which clearly reveals the logarithmic branch points at   
$\Lambda = 1/\pi$. Moreover, for $\lambda \pi <1$ we read   
off    
\begin{equation}   
{\cal E}_k= k -i\Lambda +O(\a ^2)   
\end{equation}   
while for $\lambda \pi >1$   
\begin{equation}   
{\cal E}_k=k +{1\over 2} + {i\over 2\pi }   
\ln\biggl|{1-\Lambda \pi\over 1+\Lambda \pi}\biggr|   
=k+{1\over 2} - {i\over \pi ^2\Lambda } +O(\Lambda ^{-2})
\label{energie}   
\end{equation}   
with $k$ integer. In both cases, the remaining terms, denoted by $O$,   
are purely imaginary if $\varphi=0^0$. These results were obtained in 
\cite{PT}.\\   
   
For finite $N$   
the resultant relating to the Hamiltonian $H_0- i \Lambda VV^{\dagger }$    
becomes a polynomial of order $N-1$ in $\Lambda ^2$. The complex roots which are   
the EPs are therefore not only symmetric with respect to the real axis but   
a solution $\Lambda _{{\rm EP}}$ implies also the solution $-\Lambda _{{\rm EP}}$.
As a consequence, a solution has to occur on the real $\Lambda -$axis for $N$ even.
This is a non-generic feature of the present model. For odd values of $N$
and $\varphi=0^0$, no eigenvalue is crossing an EP under variation of $\lambda$. We 
use such cases in our numerical demonstrations. \\ 
  
To illustrate how the ${\cal E}$ undergo level repulsions and how the repulsions   
are related to the  crossing points  
${\cal E}(\Lambda _{EP})$, we display in Fig.1  the crossing points and the
 eigenvalue trajectories for   
different values of $N$  and $\varphi=0^0$.  
Since the spectrum is symmetric with respect to positive and negative  
energies, only the positive part  near to the centre is drawn.
For $\Lambda = 0$ all trajectories begin at the unperturbed energies $\epsilon _k$,
move into the complex plane and then turn back  ($\epsilon _{k} \ne 0$)
 towards the real axis again.
For large $N$ the corresponding values are given by Eq.~(\ref{energie})
 (which is
valid for $N\to\infty$). With increasing $N$ the imaginary parts of the
turning and the crossing points increase, while they move nearer to each
other. The trajectory of the collective state  ($\epsilon _{k} = 0$) is
 moving on the imaginary axis
towards larger imaginary values  implying an increasing  width $\Gamma$.
We interpret the turning points as level  
repulsions of the collective state with the other levels. The descending
slope of the envelope of the turning points and the crossing points is due
to the finite size of the spectrum. \\

In Fig.2 we display the EPs in the $\Lambda -$plane for a few values of $N$. 
The sets of the EPs belonging to the same $N$ are connected by a solid line.
The zooming in of the EPs towards the accumulation point $\lambda _c = 1 / \pi$
for increasing $N$   is clearly discernible and enhanced in the insert 
of Fig.2.    
The larger $N$ the larger is the density of the EPs near to the real axis.
Additional points come in further away with each additional step of $N$. 
They are  typical edge effects and correspond to the turning points 
 of the energy trajectories at the outer edges  of the spectrum.
These points which emerge further away from the real axis quickly move in to
get near to the others while new further points come up for the following
steps of $N$.
 Eventually, in the limit
$N \to \infty$, 
all points coalesce at the accumulation point.
\\

In this way, variation of the coupling strength $\Lambda$
invokes a certain trajectory of the system 
 in the complex $\Lambda -$plane. If it hits the accumulation point
$\Lambda _c$ (or the high-density regions of EPs at finite $N$), all complex
eigenvalues coalesce (nearly coalesce) at their crossing point. This
leads to a sharp transition from a regime with $N$ resonance states 
to a regime with one collective state and $N-1$ trapped states.\\ 
 
We note that at the accumulation point $\Lambda = \Lambda _c$, which can occur only in
the limit $N\to\infty$, all infinitely many eigenstates collaps into one. It
is the global collective state. It retains its characteristics also for 
$\lambda > \lambda _c$, when the other states (the trapped states) re-emerge. This 
particular case underlines the crucial connection between the critical
point of a phase transition and the EPs.
\\

In the picket-fence model the accumulation point $\Lambda _c$
lies on the real axis. If for physical reasons,
a value $\varphi \ne 0$ has to be chosen 
with $H_0$ and $H_1$ left unchanged, a variation of 
$\lambda $ will effect a trajectory in the complex $\Lambda -$plane
that passes the high-density region (or accumulation point
for infinite $N$) at a certain distance.  This affects the sharpness 
of the transition  between the two regimes. Note that this type of
softening of the phase transition is different in nature from the one caused
by a finite value of $N$. It persists in the limit $N \to \infty$.
This underlines that $\varphi$ as well as the distribution of the EPs
determine, whether there is a phase transition in the strict thermodynamical
sense, which is associated with the sudden formation of a globally collective
state in the system. \\

As an aside we realize that the schematic model ($\varphi = 90^0 $) cannot have  
any signatures of a sharp phase transition, when the real value of $\lambda $ is 
varied from zero to large values. The accumulation point is then 
relatively far away on the imaginary axis of the coupling strength. 
Results for a variety of angles $\varphi $ but the same $H_0$ and $H_1$ have been  
investigated in \cite{PT}. \\

Previously, a criterium for a phase transition, {\it viz.}  
\begin{equation}  
B={1\over {2N+1}} \sum _{k=-N}^{N} \langle \psi _k|\psi _k\rangle     
\label{meas}  
\end{equation}  
has been introduced  in \cite{PT}. 
 Excluding the non-generic case of an accidental crossing of two (or more)
eigenvalues at an EP, $B$ is an indicator of the sharpness
of the transition. On the one hand, it is
the larger the smaller the distance between the complex eigenvalues and their
crossing points. This distance is determined by the angle $\varphi$. 
On the other hand, $B \gg 1$  only if all complex
eigenvalues reach their minimum distance simultaneously at the
same  value  $\Lambda = \Lambda _c$, i.e. if the EPs accumulate. 
In this case, $B$ shows a pronounced maximum as a function of $\Lambda$ around
$\Lambda = \Lambda _c$.\\

 In Fig.3 a few cases including the picket fence model are
illustrated. It also demonstrates a situation where a local collective
state is formed, which means that the phase transition is washed out
completely. We return to this latter aspect at the end of Section 3.3.
\\

To summarize the findings of this subsection:  the high density of  
EPs provides the mathematical mechanism for the restructuring 
of the system  under variation of the modulus $\lambda$
of the coupling parameter  $\Lambda$
towards larger values. The sharpness of the transition  which is 
evoked by the variation is determined by the distance at which the 
 high density region (accumulation point for infinite $N$) of EPs 
is passed  by the corresponding energy trajectories.  This distance is
determined by the angle $\varphi$. \\

\subsection{Random change of the unperturbed spectrum}   
   
The essential aspects of our findings   
remain unchanged if the unperturbed energies in $H_0$ and/or the   
elements of the coupling matrix $V$ deviate from the symmetrical form used   
in the previous Section.   
We address   
the question, whether and to what extent such disturbance can change the   
basic pattern, that is the formation of  a region of   high density of EPs 
or even of an accumulation point in the large $N$ limit.  
   \\

 For this purpose, we define  random perturbations in $H_0$ by
$(H_0)_{k,k}=-(N-1-2k)/2+r_k$ where the $r_k$ are random numbers from a uniform   
distribution in the interval $[-0.1,0.1]$.   
In Fig.4 
 we have drawn the EPs for the perturbed and unperturbed 
picket fence in the complex $\Lambda -$plane 
for  $N=19$. As in Fig.2, the unperturbed EPs are connected by a solid line.
The perturbed EPs do no longer lie on the   
smooth curve  but they are scattered around it.   
\\

The quantity $B$ of Eq.~(\ref{meas}) indicates the accumulative behaviour of
the EPs. It is illustrated in Fig.3 for the disturbed picket fence model
as a function of $\lambda $ for two different values of $N$ using $\varphi = 0^0$.
The excess beyond unity of $B$ around $\lambda =1/\pi$ increases  with increasing
$N$. We interpret this result  as a strong 
indication for the disturbances to be washed out, the more so the larger $N$. 
In other words, for an increasing number of states, 
the average distribution of the EPs is zooming in to $\Lambda _c
= \lambda _c = 1 / \pi$.  Hence, the EPs of the irregular system  accumulate on
the average. \\
 
Furthermore, the precice form of the distribution from which the random
changes in $H_0$ and/or $H_1$ are drawn is immaterial. In particular,
choosing the eigenvalues of $H_0$ from a Wigner (GOE) or a Poissonian
distribution, does not alter our conclusions. The existence and the position
of the accumulation point remains unaffected. \\

\subsection{Level density dependence}   
 
In real physical systems, the level density, the number and the coupling 
strength of the decay channels are in general energy dependent. 
The former is usually a monotonically increasing function, whereas 
for instance in nuclear physics the continuum coupling strength decreases.  
This energy dependence will influence the distribution of the EPs. 
\\ 
   
In \cite{PT} it has been shown that it needs a proper tuning between the 
density dependence on the one hand and the coupling dependence on the other 
in order to guarantee a phase transition at a finite value of the coupling 
strength. Given the energy dependence of the level density 
$\rho(E)$, a phase transition occurs at finite coupling strength, 
if  and only if the energy dependence 
of the coupling vector is given by the inverse function $\rho(E)^{-1}$. 
If the system obeys this condition  on the average, a phase transition 
still occurs. The critical point may  be shifted with respect to the
value of the ideal picket fence model.  
We talk about `overcompensation' when the energy dependence of the coupling    
vector   decreases (increases) at a lesser (faster) rate 
than that of the inverse behaviour of the  level density. In this case,
 numerical results \cite{PT} led to the conjecture that the critical
point is shifted to zero, i.e. $\lambda _c\to 0$ if $N\to\infty$.    
In the opposite case, which we denote as `undercompensation', a 
global collective mode and a global reorganisation of the spectrum as a whole 
is absent. Now one obtains $\lambda _c\to \infty $ if $N\to \infty $. 
The occurence of the broad mode remains then a local phenomenon
for all finite values of $\lambda$ \cite{PT}. 
A collapse of the Hilbert space as in the picket-fence model does not
occur.\\ 
   
The following substantiates the conjecture about the position of $\lambda _c$
in the under-- and overcompensated cases.
To facilitate the discussion we restrict ourselves to
 a power behaviour of the energy dependence of the coupling and 
level density. The energy dependence of the 
coupling matrix elements $|v_k|^2$ is assumed to be of the form 
$|k|^{r}$.
The  unperturbed energies (eigenvalues of $H_0$) are defined as 
$\epsilon _k = {\rm sign}(k)|k|^{t/2}$.
The secular equation Eq.~(\ref{sec1}) can be written as 
\begin{equation} 
\sum _{k=-N/2}^{N/2} {|v_k|^2\over {\cal E}-\epsilon _k}={i\over \Lambda }.
\label{sec3}
\end{equation} 
 Using (\ref{sec3}), we obtain  
\begin{equation} 
\sum _{k=-N/2}^{N/2} {|v_k|^2\over ({\cal E}-\epsilon _k)^2}=0 \label{sec4} 
\end{equation} 
which corresponds to  Eq.~(\ref{sec2}). \\

The EPs are simultaneous solutions  of Eqs.(\ref{sec3}) and
(\ref{sec4}). It is obvious from Eq.(\ref{sec4}) that the energy values at
the EP cannot be real. In fact, it is known \cite{PT},
and also discussed in connection with Fig.1,
that,  for $t =2$ and $r =0$  (picket fence), the imaginary parts of 
the energies at the EP tend to infinity for $N\to \infty $.  Numerical
evidence as well as the following consideration
supports the conjecture that this holds for arbitrary values of
$r $ and $t $ as long as $t -r >1$.  
For this purpose we rewrite Eq.(\ref{sec3}) for even $N$
\begin{equation}
2{\cal E} \sum _{k=1}^{N/2} {k^r\over {\cal E}^2-k^t}=
{i\over \Lambda }. \label{sec5}
\end{equation}
In the large $N$ limit we replace the sum in Eq.(\ref{sec5})  by
an integral and ${\cal E} $ by $iz $ and obtain
\begin{eqnarray}
-2iz \int _1^{\infty }{k^r{\rm d}k\over z^2+k^t}
&=&{-2iz\over t-r-1}F_{21}(1,1-{1+r\over t},2-{1+r\over t};-z^2)
\label{sec6}  \\
&=&{i\over \Lambda }.  \nonumber 
\end{eqnarray}
The derivative with respect to $z$ of the right hand side of Eq.({\ref{sec6}),
which is a linear combination of two Hypergeometric Functions, corresponds to
the large $N$ limit of Eq.({\ref{sec4}) (${\cal E}\to iz$).
It is straighforward to show, for instance by graphical means, that this 
derivative has no zero for finite values of $z$.
In fact, the derivative vanishes only for $z \to \infty $ irrespective
of $r$ and $t$ ($t>r+1$). We exploit this fact in Eq.(\ref{sec3}) in
that we consider the limit of the left hand side for large imaginary values
of ${\cal E} =E-i\Gamma /2$. It is found to be proportional
to $\Gamma ^{2(1+r)/t }/\Gamma$.
This yields the generalised result that the overcompensated
case occurs for $2(1+r) >t$, and the compensated and undercompensated
cases for $2(1+r) =t$ and $2(1+r) <t$, respectively \cite{PT}. In fact,
the respective limit of
the left hand side of Eq.(\ref{sec6}) is infinity, a
finite constant  and zero for $\Gamma \to \infty$.
Consequently, the corresponding values of
$\lambda $ must be zero, finite and infinity, respectively. Since these values
constitute simultaneous solutions of Eqs.(\ref{sec3}) and (\ref{sec4}), they
are the values where the EP accumulate. Furthermore it
follows, that the accumulation point is on the real axis for the compensated
case ($t =2(1+r) $).\\

To demonstrate further this result, the values of $B$ as a function of
$\Lambda = \lambda \; (\varphi =0$) are drawn for the compensated case $r=1, t=4$ in 
Fig.3. One clearly sees that the maximum of $B$ around $\lambda = 2 / \pi$
increases with $N$. As explained above, this means that, on the average, the minimum 
distance between the crossing point  and its correponding eigenvalues
decreases with increasing $N$. In the limit $N \to \infty$,
all eigenvalues will hit the crossing point. Thus, the compensated 
system behaves similar to the picket fence. Also in this case, the 
space of the eigenstates collapses at
$\Lambda = \Lambda _c = 2 / \pi$ into dimension one, and a globally collective state
is created.  For comparison, the undercompensated case with $r=0$ and $ t=4$
is drawn in Fig.3 as an example for a system which does not undergo a phase
transition. As explained above, $\lambda _c \to\infty$ in this case and $B$ has
no maximum at a finite value of $\lambda$.\\

\subsection{Phase behaviour of the wave functions} 

In Section 2 some properties of the eigenfunctions  have been pointed 
out, when a non self-adjoint Hamiltonian is considered. We resume the 
discussion in more detail here with particular emphasis on the phase of 
the wave functions when an EP is approached by the variation 
of the coupling parameter $\Lambda$. Surrounding one crossing point in 
the complex energy plane corresponds to a 
double loop around the EP in the $\Lambda -$plane,
 Eq.~(\ref{exslt}), 
since the EPs are square root branch points, Eq.~(\ref{eig2}).
We address in the following the effect of looping once around an EP
in the $\Lambda $-plane. For demonstration we use the
 simple two dimensional model introduced in Section 2, Eq.~(\ref{twod}).  \\

 The two   eigenfunctions, 
normalized according to   Eq.~(\ref{orth}), can be parametrized  
by the complex angle $\theta$: 
\begin{equation} 
\psi _1=\pmatrix {\cos \theta \cr  \sin \theta \cr },\quad 
\psi _2=\pmatrix {-\sin \theta \cr  \cos \theta \cr } 
\label{eigf} 
\end{equation}
where the  angle $\theta $ is given by 
\begin{equation} 
\tan ^2\theta ={{\cal E} _1-{\cal E} _2-(\epsilon _1-\epsilon _2)+ i\Lambda \cos 2 \omega 
\over     {\cal E} _1-{\cal E} _2+(\epsilon _1-\epsilon _2)- i\Lambda \cos 2 \omega } . 
\label{angle} 
\end{equation} 
The notation has been introduced in Eqs.~(\ref{twod}) and (\ref{eig2}). 
From this expression we read off: 
\begin{enumerate}
\item[(i)]
 at an EP (${\cal E} _1 = {\cal E} _2$) we obtain $\tan ^2 \theta =-1$ which 
implies $|\cos \theta |=|\sin \theta |=\infty $, that is the components of 
the wave functions blow up;
\item[(ii)]
 when an EP is surrounded  in the 
$\Lambda -$plane (which amounts to ${\cal E} _1-{\cal E} _2\to {\cal E} _2-{\cal E} _1$), the 
$\tan ^2\theta $ is changed into $1/\tan ^2\theta $, which corresponds to 
the change $\theta \to \theta+\pi /2$.
\end{enumerate}
 This implies 
$\psi _1\to \psi _2$ and $\psi _2\to -\psi _1$.  \\ 
 
While it may not be obvious to implement such a contour in the complex 
$\Lambda -$plane in an actual physical experiment, there could be a possibility  
to achieve the same effect by  a variation of the modulus of $\Lambda $ 
for different values of the relative coupling given by $\omega $. In fact, 
using the setting $H_0-i\lambda VV^{\dagger }$, then the EP lies 
for $\omega 
\; \raisebox{-0.4ex}{\tiny$\stackrel   {{\textstyle>}}{\sim}$}\;
45^0$ just below and for
 $\omega 
\; \raisebox{-0.4ex}{\tiny$\stackrel   {{\textstyle<}}{\sim}$}\;
 45^0$ just above 
the point $\lambda =\epsilon _2-\epsilon _1$. If we compare the two situations 
while varying $\lambda $, the two
 wave functions for $\lambda >\epsilon _2-\epsilon _1$
differ in the same way as if the EP had been surrounded. 
\\

This can be made explicit by choosing an expression for $\tan \theta $ 
which is more convenient for this purpose, {\it viz.} 
\begin{equation} 
\tan \theta = {-i\lambda \sin 2\omega 
\over {\cal E} _1-{\cal E} _2+\epsilon _1-\epsilon _2-i\lambda \cos 2\omega }. 
\label{tanc}
 \end{equation} 
The difference between the two values for $\omega $ manifests itself in the 
difference of the sign of the imaginary part of ${\cal E} _1-{\cal E} _2$, since 
different Riemann sheets have been approached. As a consequence, for large 
values of $\lambda $ the right hand side of Eq.~(\ref{tanc}) tends towards 
$\tan \omega $ for the one case and towards 
$-\cot \omega =\tan (\pi /2 + \omega )$ for the other.\\

 We suggest that using 
electro-magnetic resonators may allow control of both, the global coupling 
$\lambda $ and the relative coupling between two resonances which is given by 
$\omega $. If it should be possible to bring to interference the wave 
functions for the two different situations, that is for $\omega
\; \raisebox{-0.4ex}{\tiny$\stackrel   {{\textstyle>}}{\sim}$}\;
 45^0$ 
and for $\omega 
\; \raisebox{-0.4ex}{\tiny$\stackrel   {{\textstyle<}}{\sim}$}\;
 45^0$, the different phases are expected to be 
observable. \\

The situation described is reminiscent of Berry's phase \cite{Berry}.
However, we stress that in our discussion we are dealing with non self-adjoint
operators, and an EP is therefore not to be confused with
Berry's diabolic points.  This difference was pointed out in 
\cite{hemond}.
In fact, a generic EP (i.e. when just two levels coalesce)
gives rise to a double pole in the corresponding Green's function or S-matrix.
 The mechanism for this 
to occur is again related to the vanishing of the norm of the eigenvectors
when $\Lambda \to \Lambda _{{\rm EP}}$. The double pole is an additional signature for 
an EP. It is different from the simple pole emerging in the Green's
function (S-matrix), if a usual single resonance or an incidental
degeneracy of two resonances occurs, say, owing to some symmetry. The latter
situation is, however, an unlikely event as it
needs the tuning of four parameters (six parameters in the case of no 
time reversal) to achieve it. In contrast, the existence of the EPs
 is an intrinsic mathematical feature.

\section{Summary and Outlook}   
   
In this paper, we studied the relation between phase transitions, collective
states and the distribution of EPs of open quantum systems. Since both the
collectivity and the phase transition
are well pronounced in the one-channel case, we restricted ourselves 
to  an $N-$dimensional quantum system
coupled to one open decay channel. Our results are as follows:\\

(i)  A necessary condition for a phase transition to occur is that 
 the EPs accumulate   in the complex $\Lambda -$plane. This result has been
shown analytically for the picket-fence model  and some generalisations,
and is confirmed by numerical results using the quantity $B$.
  The quantity $B$ 
is a measure for the average minimum distance of the 
eigenvalues to its crossing point in the complex  energy plane.
In the cases considered, $B$ has a pronounced maximum at the critical value
$\Lambda _c$ at which the phase transition occurs. The increase of $B$
with the number $N$ of states  is a further indication that the 
 EPs accumulate in the complex $\Lambda -$plane.
\\

(ii)
The choice of the angle $\varphi$ 
 is dictated by the physical situation.
Only if it is equal to the phase of $\Lambda _c$, the system
hits the accumulation point (or goes through the region of high 
density of EPs), when $\lambda$ is varied. Otherwise,  
the system may pass the vicinity of the accumulation point 
(or the region of high density of EPs) and a phase transition 
in the strict thermodynamical sense does not occur. 
In all cases considered, we find that $\varphi$  has to be $ 0^0$
for hitting the accumulation point $\Lambda _c$
and for a genuine phase transition to emerge.
\\

(iii)
If the system hits the accumulation point, the infinite--dimensional space
collapses to a one-dimensional function space, since the infinitely many
eigenfunctions are identical at $\Lambda = \Lambda _c$. This is in contrast to the
function space of a usual $N-$fold degeneracy, where $N$ independent
eigenfunctions occur. The eigenfunction of the collective state
contains contributions of all basis functions of the unperturbed system $H_0$.
It is a globally collective state which is created by the system as a whole.\\

(iv) If the system does not hit the accumulation point,  
the formation of the collective state is not
related to  a genuine phase transition. This is
in particular the case for a
system with a purely hermitian coupling matrix $H_1$
($\varphi = 90^0$) which creates a state with large internal 
collectivity. 
\\

(v)
If the accumulation point is at $\Lambda _c \to \infty$, the 
formation of the collective state 
occurs successively and  locally. Its wavefunction  contains 
contributions only from those resonance states which are  overlapped by it.
\\

 Exploiting the fact that the crossing points in the complex energy plane 
 are  square root branch points, we suggest a possibility for
an experimental study
of the level repulsion of the complex eigenvalues. 
For this, the relative coupling
strength  of the states to the continuum as well as
 the overall value of their coupling strength should be controllable. 
A signature of local resonance crossing is given by a particular change of
the phase of the complex eigenfunctions. Work towards this aim is in progress.
\\[.3cm]
   
\noindent
{\bf Acknowledgement:} Valuable discussions with T.~Gorin, C.~Jung,
S.~Muraviev and
G.~Soff are gratefully acknowledged.
The present investigations are supported by  DFG
and SMWK.\\

\vspace{1cm}

\noindent
{\Large\bf Figure Captions}

\vspace{1.5cm}

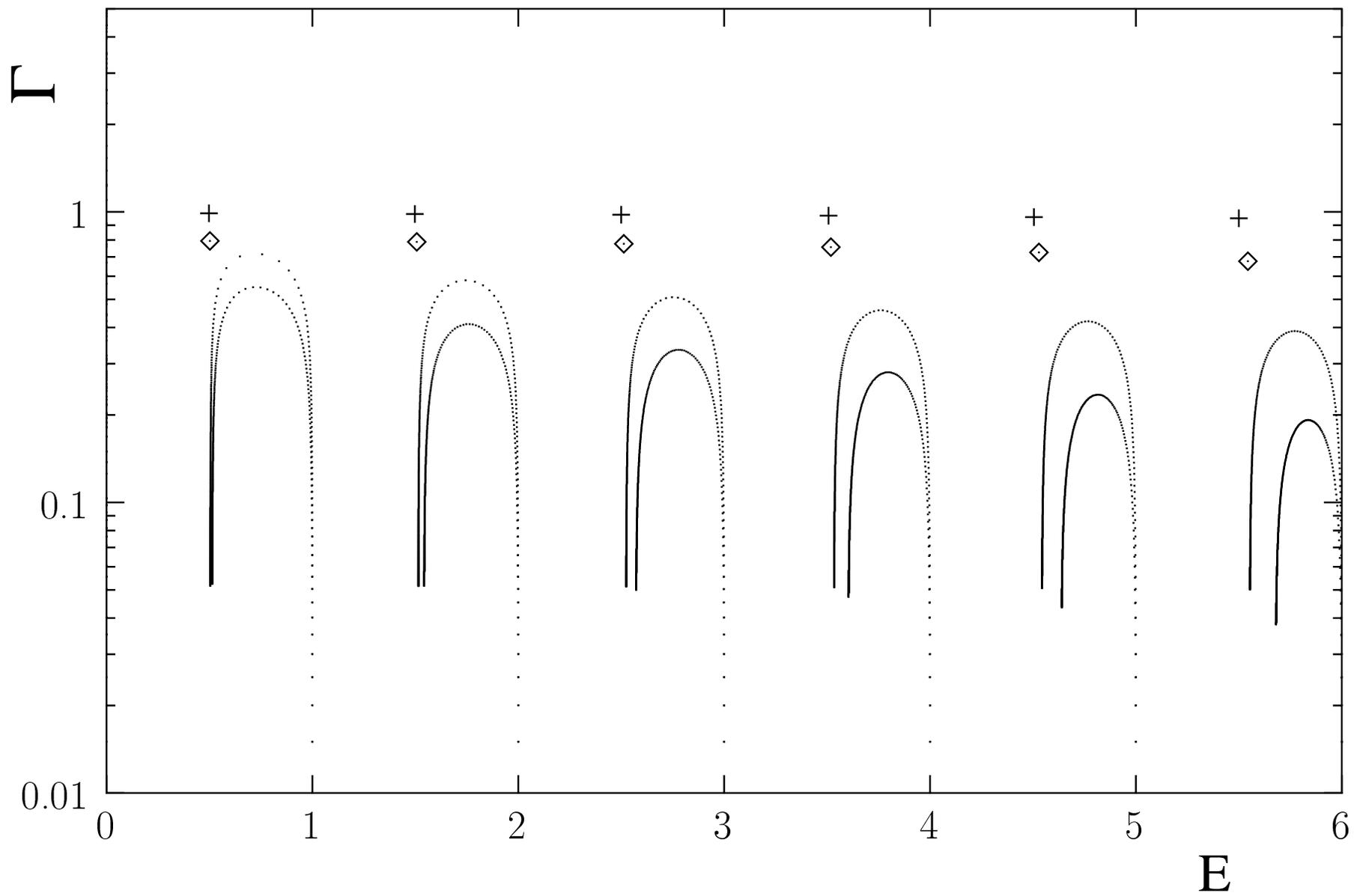
{\bf Figure 1}\\

\noindent
The trajectories of the eigenvalues for $N = 15$ and $N=43$ for increasing 
$\lambda \in [0.001,2]$  in steps of 0.001.
The crossing points are marked by diamonds ($N=15$) and crosses
 ($N=43$). 
\\[.2cm]

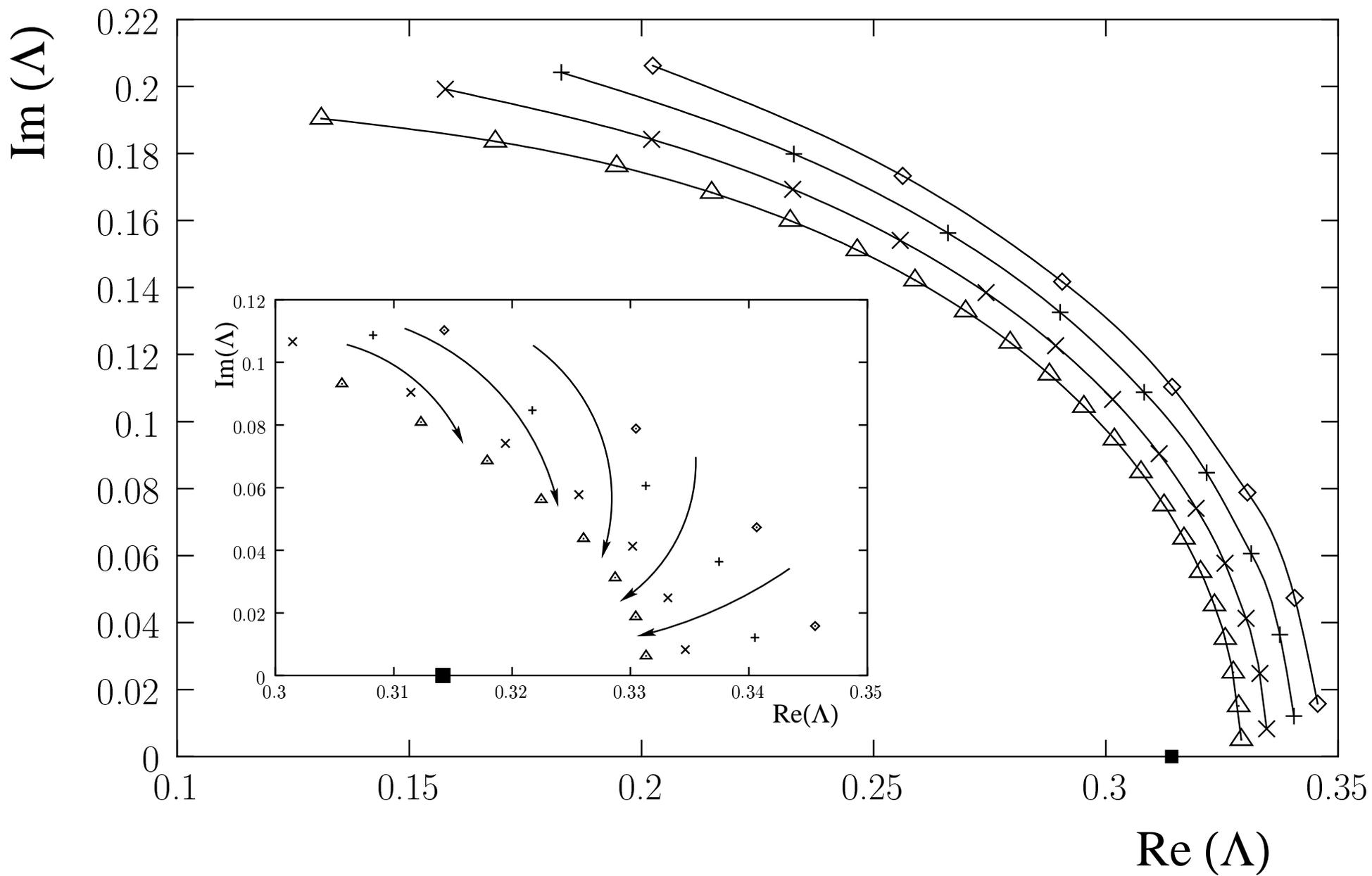
{\bf Figure 2}\\

\noindent
The exceptional points in the complex $\Lambda -$plane for 
$N=15$ (rhombs), $N=19$ (plus signs), $N=27$ (crosses) and
$N=43$ (triangles). The inset is a magnification around the accumulation
point $\Lambda _c$ (black square).
The arrows indicate the changes of the EPs with increasing $N$.
\\[.2cm]

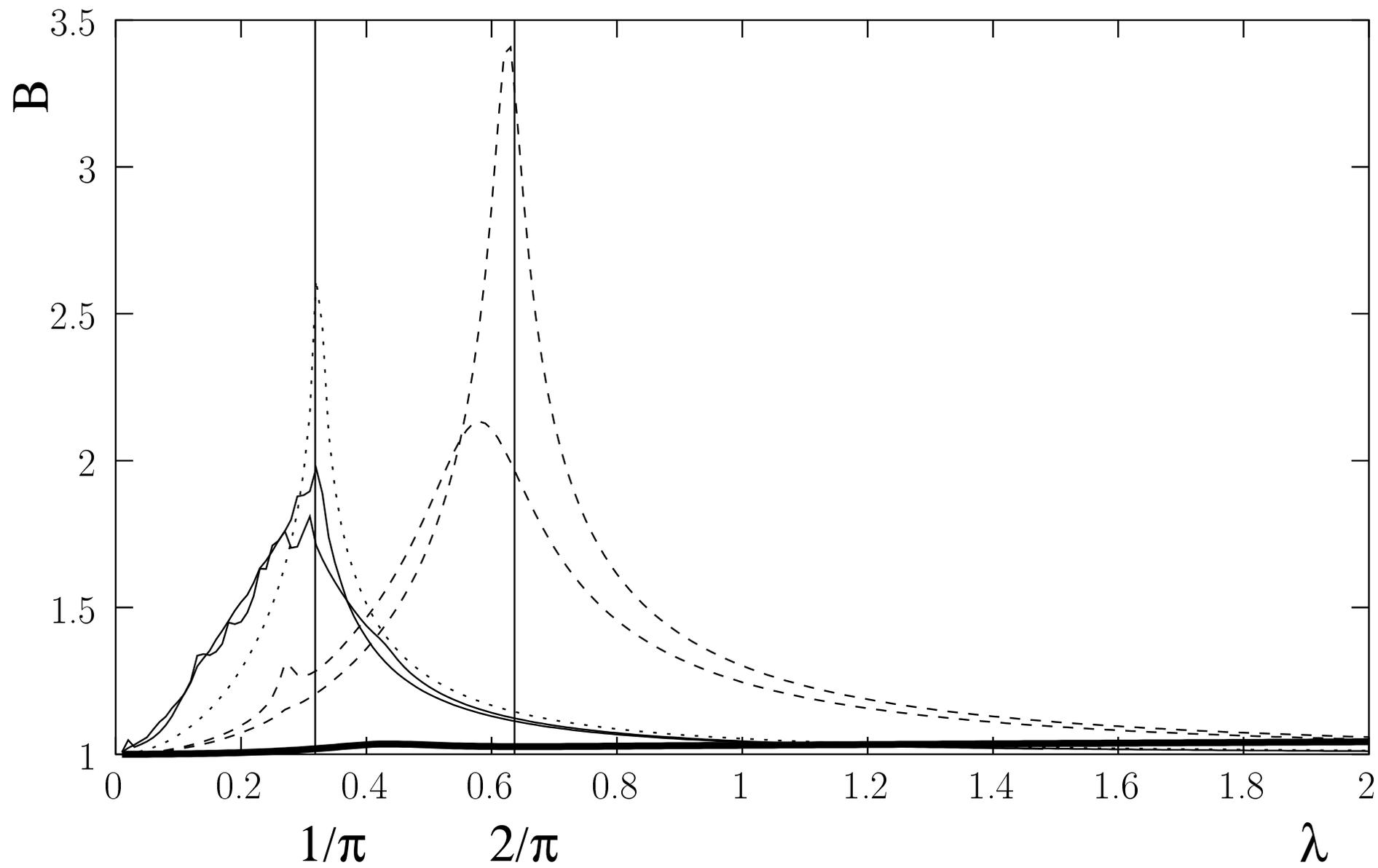
{\bf Figure 3}\\

\noindent
$B$ as a function of $\lambda $ for different systems with $\varphi = 0$:
ideal picket fence with $N=101$ (dotted line), randomly perturbed picket fence
with $N=101, \; 1001$ (solid lines), compensated case with 
$r=1,\; t=4$ and $N=101, \; 1001$ (dashed lines), undercompensated case
with $r=0, \; t=4$ and $N=101$ (thick line).
The two values $\lambda _c = 1 / \pi $ and $ 2/\pi$, referring to $N\to \infty $, 
are indicated by a vertical solid line.
\\[.2cm]

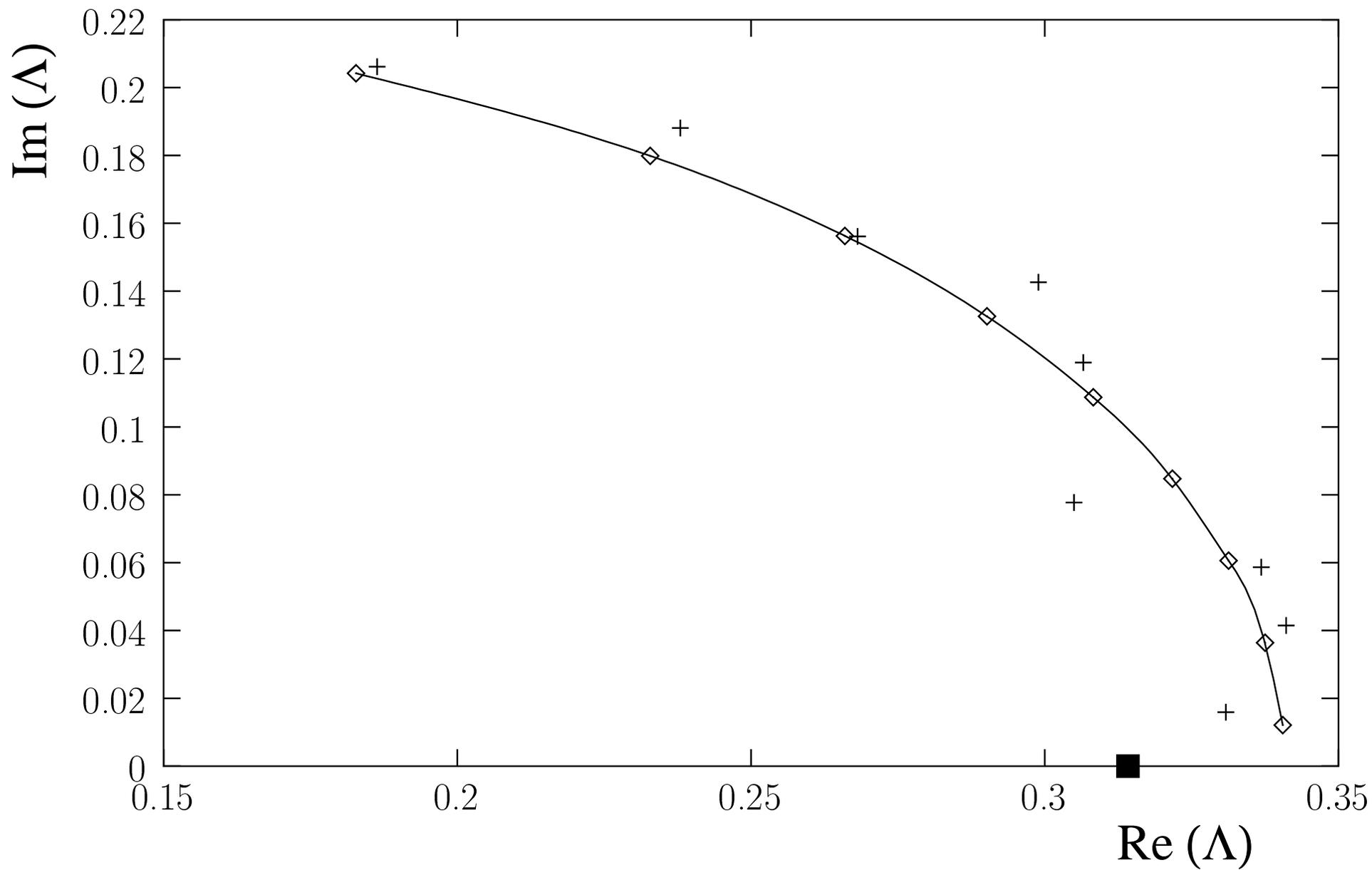
{\bf Figure 4}\\

\noindent
The EPs for $N=19$ for the ideal picket fence (rhombs) and the randomly
perturbed picket fence (plus signs).
$\Lambda _c$ is denoted by a black square.
\\[.2cm]


\begin{thebibliography}{99}   

\bibitem{trap} I.~Rotter, Rep. Prog. Phys {\bf 54}, 635 (1991);
V.V.~Sokolov and V.G.~Zelevinsky, Ann.~Phys.~(N.Y.) {\bf 216}, 323 (1992);
M.~Desouter-Lecomte, J.~Li\'evin and V.~Brems, J.~Chem.~Phys.~{\bf 103},
15 (1995) 

\bibitem{dirose} F.M.~Dittes, I.~Rotter and T.H.~Seligman, 
Phys.~Lett.~{\bf A158}, 14  (1991)

\bibitem{mudiisro} M.~M\"uller, F.-M.~Dittes,  W.~Iskra and I.~Rotter, 
Phys.~Rev.~{\bf E52} 5961 (1995)

\bibitem{sorosamu}
V.V.~Sokolov, I.~Rotter, D.V.~Savin and M.~M\"uller, Phys.~Rev.~{\bf C56},
1031 and 1044 (1997)

\bibitem{godimurose} T.~Gorin, F.M.~Dittes, M.~M\"uller, I.~Rotter and
T.H.~Seligman, Phys.~Rev.~{\bf E56}, 2481  (1997)

\bibitem{pegoro}
E.~Persson, T.~Gorin and I.~Rotter, Phys.~Rev.~E  (in press)

\bibitem{PT} C. Jung, M.~M\"uller and I.~Rotter, quant-ph/9804020  

\bibitem{schemat} G.~Brown and M.~Bolsterli, Phys.~Rev.~Lett.~{\bf 3}, 472
 (1959)

\bibitem{hesa} W.D. Heiss and A.L. Sannino, J. Phys. {\bf A23} 1167 (1990);   
W.D. Heiss and A.L. Sannino, Phys. Rev. {\bf A43} 4159 (1991)   

\bibitem{wdh} W.D. Heiss, Phys. Rep. {\bf 242}, 443 (1994)   

\bibitem{kohe} W.D. Heiss and A.A. Kotz\'e, Phys. Rev. {\bf A44}, 2403   
(1991); A.A. Kotz\'e and W.D. Heiss, J. Phys. A: Math Gen, {\bf 27}, 3059   
(1994)   

\bibitem{Berry} M.V.~Berry, {\it Quantum Chaos}, ed.~by G.~Casati (London:
Plenum) 1985; Proc.~R.~Soc.~{\bf A239}, 45 (1983)

   
\bibitem{hemond} E. Hernandez and A. Mondragon, Phys.Lett {\bf B326},  
1, (1994); A. Mondragon and E. Hernandez, J. Phys.~{\bf A26}, 5595 (1993) 
   
   
\end{thebibliography}
\end{document}